\documentclass[aps,pre,reprint,superscriptaddress]{revtex4-1}

\usepackage[dvipdfmx]{graphicx}
\usepackage{dcolumn}
\usepackage{bm}
\usepackage[usenames]{color}

\begin{document}

\title{Princess and the Pea at the nanoscale: Wrinkling and delamination of graphene on nanoparticles}
\author{Mahito Yamamoto}
\affiliation{Materials Research Science and Engineering Center, Department of Physics, University of Maryland, College Park, MD 20742, USA}
\affiliation{Center for Nanophysics and Advanced Materials, Department of Physics, University of Maryland, College Park, MD 20742, USA}

\author{Olivier Pierre-Louis}
\affiliation{Institut Lumi\`{e}re Mati\`{e}re, UMR5306 Universit\'{e} Lyon 1-CNRS, Universit\'{e} de Lyon 69622 Villeurbanne cedex, France}

\author{Jia Huang}
\affiliation{Center for Nanophysics and Advanced Materials, Department of Physics, University of Maryland, College Park, MD 20742, USA}

\author{Michael~S. Fuhrer}
\affiliation{Materials Research Science and Engineering Center, Department of Physics, University of Maryland, College Park, MD 20742, USA}
\affiliation{Center for Nanophysics and Advanced Materials, Department of Physics, University of Maryland, College Park, MD 20742, USA}

\author{Theodore~L. Einstein}
\affiliation{Materials Research Science and Engineering Center, Department of Physics, University of Maryland, College Park, MD 20742, USA}
\affiliation{Center for Nanophysics and Advanced Materials, Department of Physics, University of Maryland, College Park, MD 20742, USA}

\author{William~G. Cullen}
\affiliation{Materials Research Science and Engineering Center, Department of Physics, University of Maryland, College Park, MD 20742, USA}
\affiliation{Center for Nanophysics and Advanced Materials, Department of Physics, University of Maryland, College Park, MD 20742, USA}

\date{\today}


\begin{abstract}
Thin membranes exhibit complex responses to external forces or geometrical constraints. A familiar example is the wrinkling, exhibited by human skin, plant leaves, and fabrics, resulting from the relative ease of bending versus stretching. Here, we study the wrinkling of graphene, the thinnest and stiffest known membrane, deposited on a silica substrate decorated with silica nanoparticles. At small nanoparticle density monolayer graphene adheres to the substrate, detached only in small regions around the nanoparticles. With increasing nanoparticle density, we observe the formation of wrinkles which connect nanoparticles. Above a critical nanoparticle density, the wrinkles form a percolating network through the sample. As the graphene membrane is made thicker, global delamination from the substrate is observed. The observations can be well understood within a continuum elastic model and have important implications for strain-engineering the electronic properties of graphene.
\end{abstract}

\maketitle

\section{Introduction}

Thin films supported on substrates are of technological importance and are commonplace in biological systems such as cell walls and hard skins on soft plant and animal tissues. As the thickness $t$ of a slab of material is reduced, it becomes more susceptible to out-of-plane deformation (bending) compared to in-plane deformation (stretching), resulting in morphological transitions. For example, thin films deposited on soft compliant substrates display wrinkling patterns under compressive stress \cite{Bowden}, and membranes resting on fluids wrinkle by capillary forces \cite{Huang1}. This wrinkling is a ubiquitous phenomenon found in systems ranging from human skin to draping fabric \cite{Cerda1, Cerda2, Witten} and has also been exploited to fabricate flexible electronic devices \cite{Khang, Sun}.

Due to the inextensible but bendable nature of the \textit{sp}$^2$ carbon bond, graphene's effective mechanical thickness $t_{\rm{eff}} = (12\kappa/E_{2D})^{1/2}$ is less than 1 {\AA} \cite{Huang2}, where $\kappa \approx$ 1 eV \cite{Fasolino} is the bending rigidity and $E_{2D} \approx 2.12 \times 10^3$ eV/nm$^2$ \cite{Lee} is the tensile rigidity; this is likely the smallest mechanical thickness ever achieved for any material. Graphene has been anticipated to exhibit a rich variety of wrinkling and delamination behaviors \cite{Wang, Li1, Aitken, Wagner, Pierre-Louis, Kusminskiy}. Although graphene adheres conformally to smooth nano-scale features with high fidelity \cite{Cullen}, graphene wrinkling has been observed under compressive \cite{Bao} or tensile \cite{Li2} stress caused by thermal cycling, and graphene on periodically corrugated elastic \cite{Scharfenberg1} and metal \cite{Scharfenberg2} substrates exhibit transitions from adhesion to delamination. Morphological features such as wrinkles and conical singularities may produce non-uniform strain in graphene \cite{Pereira1}, which produce both scalar and vector potentials and mimic the effect of a magnetic field on graphene's electronic structure \cite{Castro Neto}. Recent experimental results suggest these effective ``pseudomagnetic'' fields can exceed several hundred Tesla \cite{Levy}, and graphene devices based on ``strain engineering'' have been proposed \cite{Guinea, Pereira2}. Understanding the mechanical response of graphene to non-uniform stress is thus a critical first step toward strain engineering its electronic properties.

Here we report a systematic study of the wrinkling of graphene membranes supported on SiO$_2$ substrates with randomly placed topographic perturbations, produced by SiO$_2$ nanoparticles. The wrinkling is probed as a function of nanoparticle density $\rho_{np}$ and membrane thickness (using multilayer graphene). At low $\rho_{np}$, monolayer graphene largely conforms to the substrate except for small regions around the nanoparticles, where graphene is detached. Wrinkles (or folds or ridges) form as $\rho_{np}$ increases, connecting pairs of protrusions. The observed maximum wrinkle length is predicted quantitatively within a simple elastic model. Above a critical density the wrinkles percolate to form a network spanning the entire sample. As the thickness of graphene increases, it stiffens and delaminates instead of wrinkling. Since the wrinkling acts to remove inhomogeneous in-plane elastic strains through out-of-plane buckling, our results can be used to place limits on the possible in-plane strain magnitudes that may be created in graphene to realize strain-engineered electronic structures \cite{Pereira1, Castro Neto, Levy, Guinea, Pereira2}.

\begin{figure*}[t]
\includegraphics[width=170 mm]{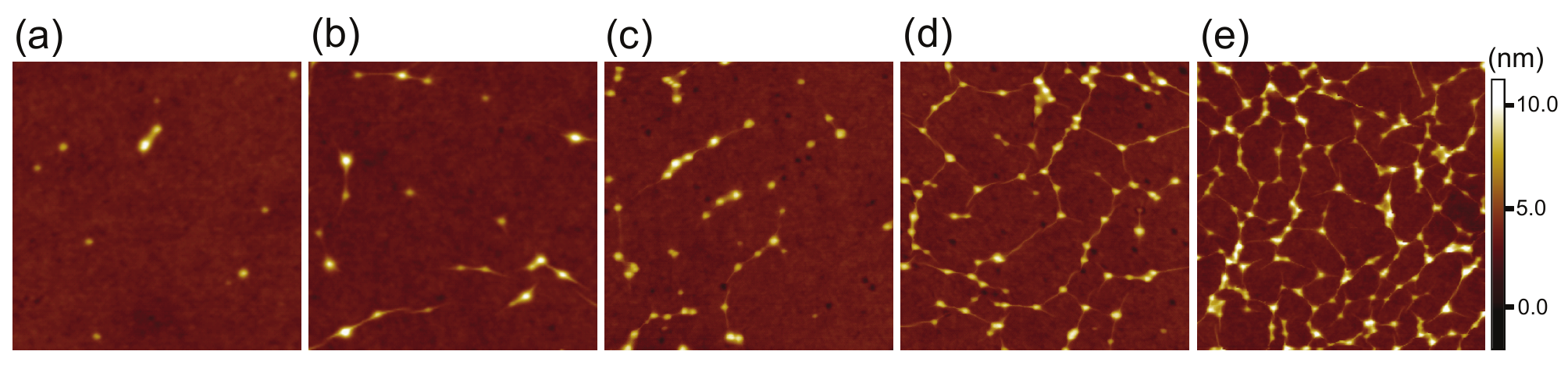}
\caption{AFM images (1 $\times$ 1 $\mu$m$^2$) of graphene on SiO$_2$ nanoparticle/SiO$_2$ substrates for a nanoparticle density of (a) 11, (b) 22, (c) 49, (d) 90, and (e) 170 $\mu$m$^{-2}$, respectively. \label{fig.1}}
\end{figure*}

\section{Experimental details}

Silica-nanoparticle colloidal dispersions (Nissan Chemical America Corp., Snowtex-O) were diluted to various concentrations of 0.5$-$3.0 wt\% by deionized water (Fisher Scientific, Water HPLC Grade). The diluted suspensions were sonicated for 30 min in a water bath to break agglomerations before spin-coating the nanoparticles onto a substrate. Spin-coating was performed on Si substrates with a 300 nm-thick oxide layer at 4000 rpm for 30 seconds. The density of nanoparticles on substrates ranged from 2 $\mu$m$^{-2}$ to 258 $\mu$m$^{-2}$, depending on the concentrations of the nanoparticle dispersions. After spin-coating, the samples were completely dried on a hotplate at $\sim$ 150 ${}^\circ$C for 2 hours. Graphene flakes were mechanically exfoliated from Kish graphite onto SiO$_2$ substrates covered with the silica nanoparticles (mean diameter 7.4 $\pm$ 2.2 nm; Ref.~\onlinecite{Supplementary}). Thicknesses of graphene films were identified with an optical microscope, atomic force microscopy (AFM), and/or Raman spectroscopy. The sizes of graphene sheets were typically more than 10 $\mu$m $\times$ 10 $\mu$m, which were much larger than an estimated distance between nanoparticles of $\sim$ 700 nm at the smallest nanoparticle density of 2 $\mu$m$^{-2}$. Thus, we rule out the possibility of finite size effects in the following analyses. The samples were introduced into a vacuum chamber with a base pressure of $\sim$ 10$^{-7}$ Torr and annealed at $\sim$ 500 ${}^\circ$C for $\sim$ 5 hours in order to remove any adhesive tape residue and to achieve equilibrium structures. After the annealing procedure, we observed surfaces of graphene flakes of various thicknesses in air using AFM in tapping mode with silicon cantilevers with a nominal tip radius of $<$ 10 nm (Olympus, OMCL-AC160TS).

\section{Results and Discussion}

\subsection{Wrinkling of monolayer graphene}

\begin{figure}[h]
\includegraphics[width=85 mm]{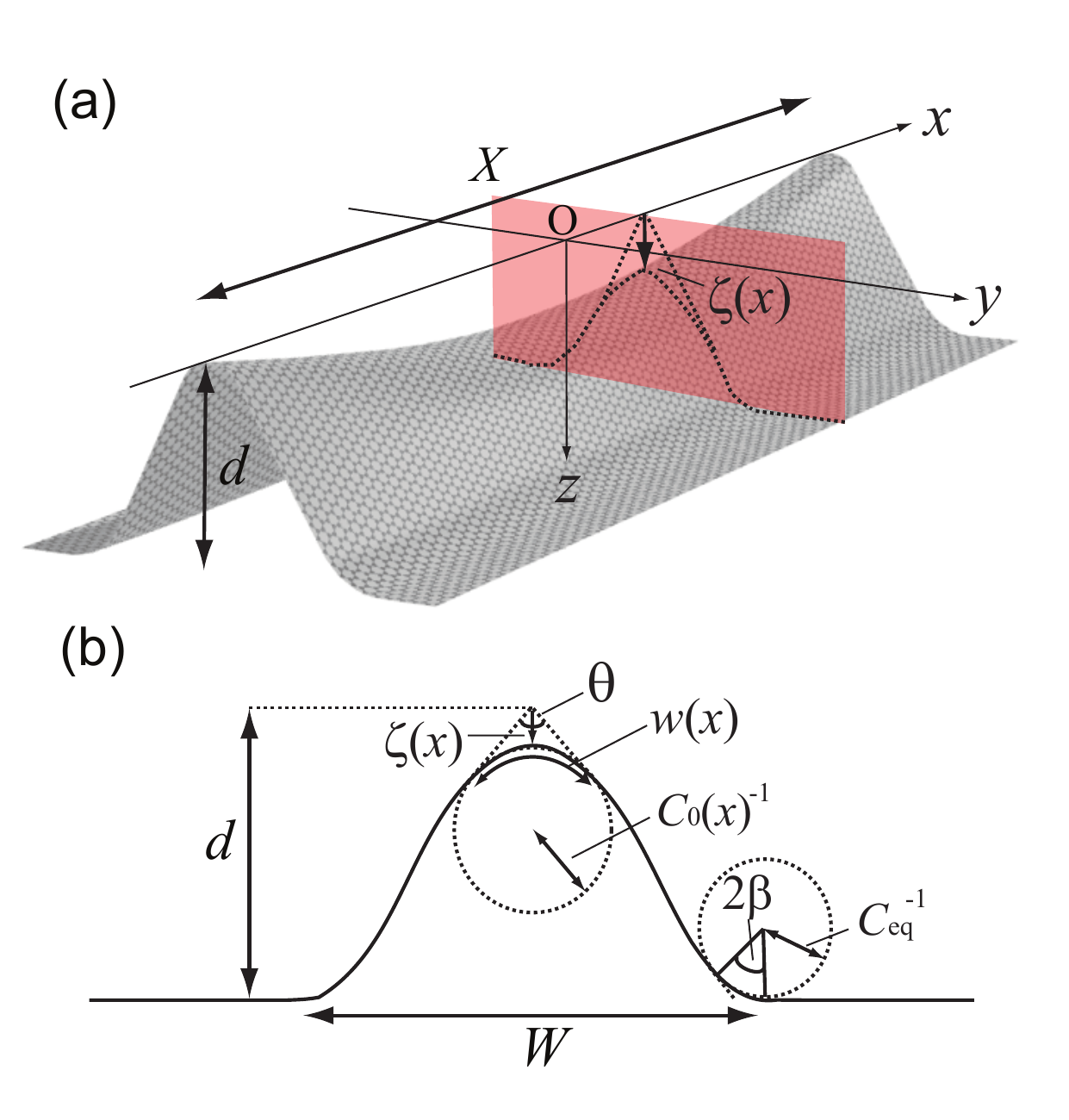}
\caption{Schematics of (a) a wrinkle formed between two nanoparticles with diameters $d$ and (b) the wrinkle profile along the transverse direction as represented by shaded area in (a). \label{fig.2}}
\end{figure}

Figure~\ref{fig.1} shows typical AFM images of monolayer graphene supported on nanoparticles for various densities $\rho_{np}$. At $\rho_{np} = 11$ $\mu$m$^{-2}$ (Fig.~\ref{fig.1}a), graphene adheres conformally to the substrate, as noted previously \cite{Cullen, Ishigami, Lui, Xue}, with predominantly isolated protrusions at the nanoparticle locations. At $\rho_{np} = 22$ $\mu$m$^{-2}$ (Fig.~\ref{fig.1}b), some nanoparticle-induced protrusions are linked by wrinkles (we use the term ``wrinkle'' in accordance with the literature on graphene). Additional wrinkles with one free termination are also observed. The wrinkles between protrusions sag. If the protrusions have comparable heights as we assume below for simplicity, the wrinkle sags in the middle, while if the height difference is large, the wrinkle sags asymmetrically toward the protrusion of smaller height. With further increase in nanoparticle density, the wrinkles connecting the protrusions proliferate (Fig.~\ref{fig.1}c), and ultimately a wrinkle network spans the sample (Figs.~\ref{fig.1}d and e). These observations indicate the presence of a critical distance $X_c$ between nanoparticles, below which wrinkling is induced.

We now analyze the critical nanoparticle separation $X_c$. The ridge running along the wrinkle between two nanoparticles with diameters $d$ separated by $X$ follows a catenary-like profile with a deflection $\zeta (x)$ as shown in Fig.~\ref{fig.2}a. Additionally, as represented in Fig.~\ref{fig.2}b, the profile of the ridge along the transverse ($y$) direction can be characterized with the dihedral angle $\theta$ and the curvature radius $C_0(x)^{-1}$. The contour of the wrinkle results from the balance between elasticity and adhesion. A wrinkle of length $X$ much larger than its average width $W_{\rm{avg}}$ costs adhesion energy $\sim \Gamma XW_{\rm{avg}}$, where $\Gamma$ is the graphene-SiO$_2$ adhesion energy per area. The deflection of elastic sheets between two protrusions creates a stretching strain $\epsilon \sim (\zeta_0/W_{\rm{avg}})^2$ along the crease of the ridge in a region of width $\sim \zeta_0$ \cite{Lobkovsky}, where $\zeta_0 \equiv \zeta(0)$ is the maximum deflection. Balancing the stretching energy $\sim E_{2D}X\zeta_0\epsilon^2$ with the adhesion energy, and assuming that $W_{\rm{avg}} \sim d$, one finds that $\zeta_0 \sim X^{4/5}d^{1/5}(\Gamma/E_{2D})^{1/5}$. Precluding wrinkles with deflections larger than $d$, one finds a maximum wrinkle $X_c \sim d(E_{2D}/\Gamma)^{1/4}$.

We next present a detailed elastic analysis of the wrinkle shape including bending energy, which provides an expression for the full deflection profile $\zeta(x)$ and recovers the scaling law for $X_c$. Assuming the opening angle $\theta$ is independent of $x$ as validated in Ref.~\onlinecite{Lobkovsky}, we find the width of the deformed region $w(x) = (\pi - \theta)C_0(x)^{-1}$ and the deflection $\zeta(x) = [1/\sin(\theta/2) - 1]C_0(x)^{-1}$ within the effective one-dimensional model. Furthermore, we assume that the stretching strain in the $y$ direction is irrelevant according to Ref.~\onlinecite{Lobkovsky}. Then, the stretching strain is also given in one dimension (in the $x$ direction) by $\epsilon_x = [1 + (\partial_x\zeta)^2]^{1/2} -1 \approx (\partial_x\zeta)^2/2$. We find the stretching energy $E_s$ and the bending energy $E_b$:

\begin{equation}
E_s=\frac{E_{2D}}{2}\int dx\, w(x)\epsilon_x^2\label{eq.1}
\end{equation}

\noindent 

\begin{equation}
E_b=\frac{\kappa}{2}\int dx\, w(x)C_0(x)^2.\label{eq.2}
\end{equation}
\noindent The adhesion energy cost is proportional to the area of the substrate uncovered by the membrane:

\begin{eqnarray}
E_{\Gamma}=\Gamma\int dx\, W,\label{eq.3}
\end{eqnarray}
where $W$ is the base of the wrinkle profile as illustrated in Fig.~\ref{fig.2}b.
In addition, we take into account bending and adhesion at the foot of the wrinkle, which cost bending energy $E_{b'}$ and adhesion energy $E_{\Gamma '}$ (see Appendix A for details), but these turn out to be negligible.

\begin{figure}[t]
\includegraphics[width=75 mm]{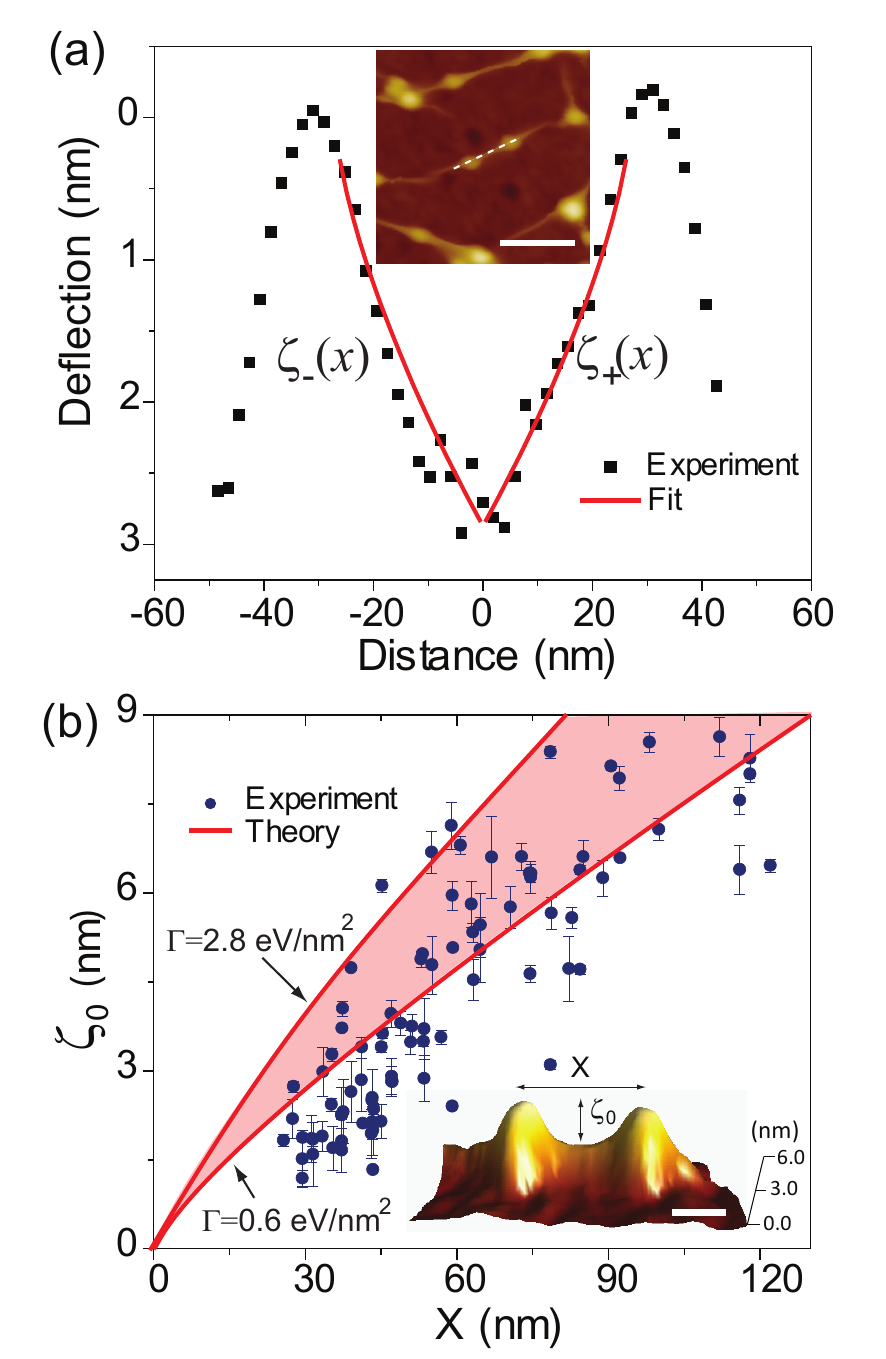}
\caption{(a) Profile of the wrinkle along the white dotted line in the AFM image shown in the inset. The scale bar in the inset is 50 nm. The solid red lines are theoretical expectations. (b) The maximum deflection $\zeta_0$ as a function of the wrinkle length. The error bar indicates the uncertainty of $\zeta_0$ due to the height difference between the protrusions. We choose wrinkles formed between protrusions where the height differences are less than 10 $\%$. The area shaded in red is the theoretical prediction for scaling of $\zeta_0$ with $X$ for $\Gamma = 0.6-2.8$ eV/nm$^2$. The inset is a typical AFM image of the wrinkles formed between the two protrusions. The scale bar is 20 nm. \label{fig.3}}
\end{figure}



At equilibrium, we expect $\delta E_{\rm{tot}}/\delta\zeta = 0$, where $E_{\rm{tot}} = E_s + E_b + E_{\Gamma}+E_{b'}+E_{\Gamma '}$, leading to a differential equation for deflection $\zeta$. By solving the differential equation under two boundary conditions $\zeta(\pm X/2) = 0$, we find the deflection on both sides of the center of the wrinkle (see Appendix A):
\begin{equation}
\hspace{-0.01mm}\zeta_{\pm}(x)=\left(\frac{27\kappa}{4E_{2D}}\right)^{1/6}\!
\left[\frac{1}{\sin (\theta/2)}\! -\! 1\right]^{1/3}\!\left(\frac{X}{2}\! \mp\! x\right)^{2/3}\! .\label{eq.7}
\end{equation}
In Fig.~\ref{fig.3}a, we show the line profile along a wrinkle formed between two protrusions. The observed deflection is well fitted by the theoretical prediction $\zeta_{\pm}(x)\sim(X/2 \mp x)^{2/3}$ with a prefactor of 0.32 nm$^{1/3}$. Then, using $d = 7.4 \pm 2.2$ nm, $E_{2D} = 2.12 \times 10^3$ eV/nm$^2$ \cite{Lee}, $\kappa = 1$ eV \cite{Fasolino}, and $\Gamma =0.6-2.8$ eV/nm$^2$ \cite{Cullen, Ishigami, Koenig}, we minimize the total energy $E_{\rm{tot}}$ numerically with respect to $\theta$ for given $X$. Using Eq.~\ref{eq.7}, we find the maximum deflection $\zeta_0 \equiv \zeta (0)$ as a function of $X$ as illustrated in Fig.~\ref{fig.3}b. The maximum deflection $\zeta_0$ monotonically increases with $X$, in good agreement with the observations. The theoretical model for a deflection is based on the assumption that a wrinkle is formed between two sharp peaks. The finite sizes of the protrusions may be a cause of decrease of the deflection below the theoretically expected range in Fig.~\ref{fig.3}b. Furthermore. we attribute the most likely source of uncertainty to the observed dispersion in nanoparticle sizes.

Since a wrinkle is geometrically suppressed if $\zeta (0) > d$, the maximum length of the wrinkle is determined by a condition that $\zeta (0) = d$. From Eq.~\ref{eq.7}, we find the maximum length $X_c = 104-65$ nm along with 
$\theta = 35{}^\circ-14{}^\circ$ for the adhesion energy $\Gamma = 0.6-2.8$ eV/nm$^2$, respectively, in rough agreement with the observed maximum wrinkle length of $\sim 200$ nm \cite{Supplementary}. Neglecting the contributions of $E_{b'}$ and $E_{\Gamma '}$, $X_c$ slightly decreases to $96-62$ nm. The discrepancy between the theoretical predictions and the observations is likely due to the fluctuations in the nanoparticle sizes $d$, which strongly influence the wrinkle length $X_c$ ($X_c$ increases in proportion to $d$). In Appendix B, we discuss the link between the detailed elastic model and the scaling analysis.

\subsection{Random wrinkling model}

\begin{figure}[t]
\includegraphics[width=75 mm]{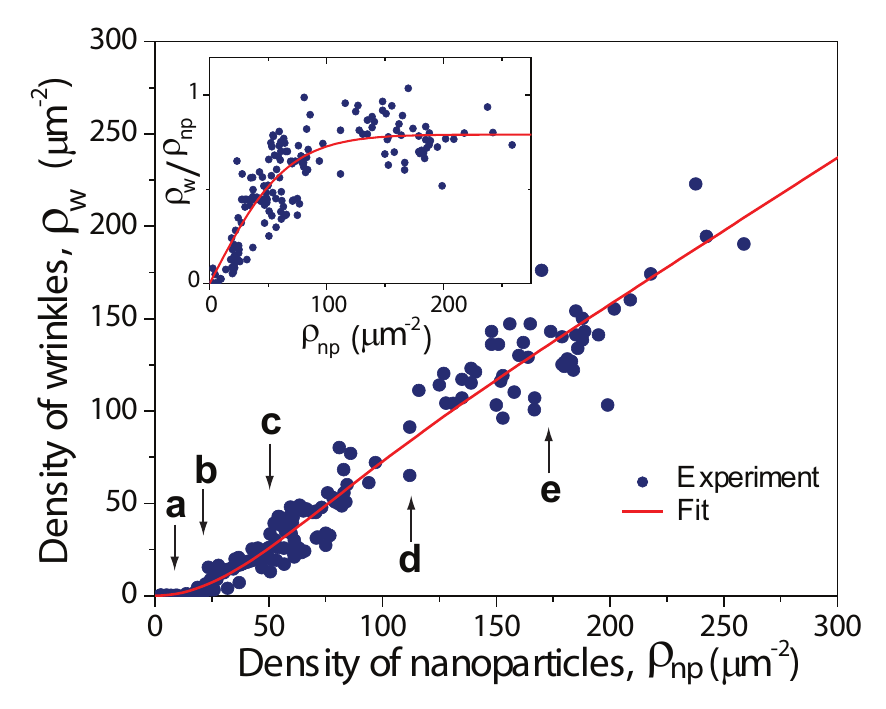}
\caption{The density of wrinkles $\rho_w$ and the mean number of wrinkles per protrusion $\rho_{w}/\rho_{np}$ (inset) as functions of nanoparticle density $\rho_{np}$. Each arrow corresponds to the AFM images shown in Figs.~1a-e. The solid red lines are fits described in text.\label{fig.4}}
\end{figure}

\begin{figure*}[t]
\includegraphics[width=170 mm]{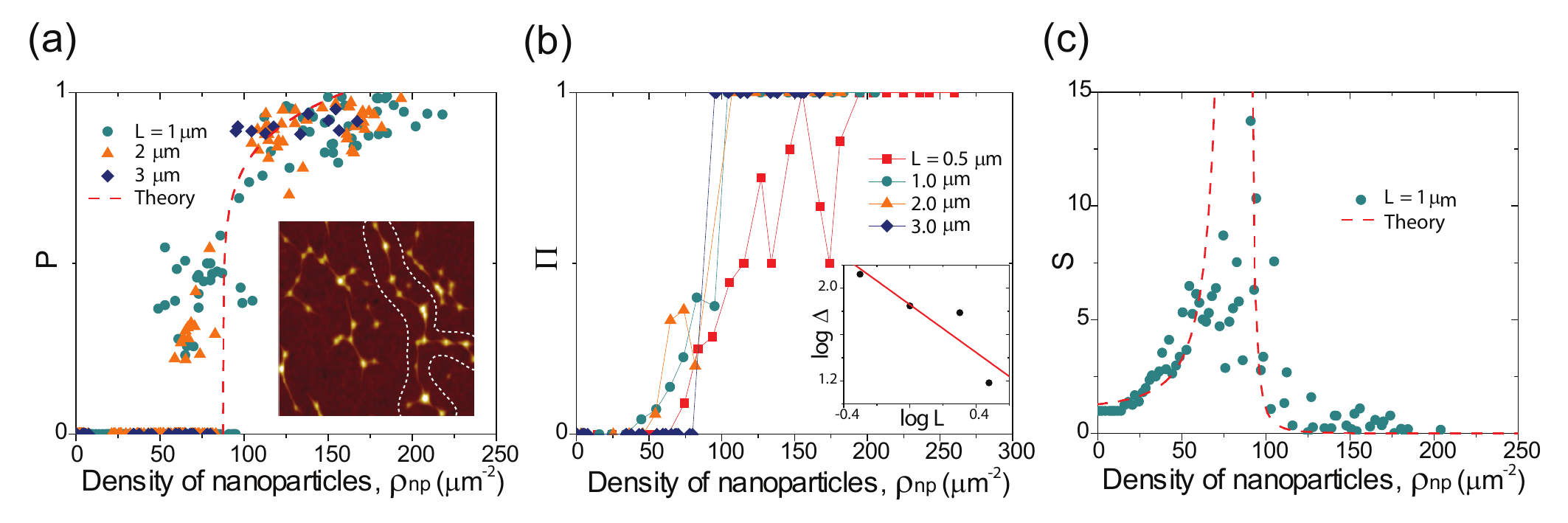}
\caption{(a) $P$ as a function of nanoparticle density for $L =$ 1, 2, and 3 $\mu$m. The inset is an AFM image (1 $\times$ 1 $\mu$m$^2$) of graphene on the SiO$_2$ nanoparticles with a density of 57 $\mu$m$^2$, showing the percolating cluster highlighted by the dashed curve. (b) $\Pi$ as a function of the density of nanoparticles for $L =$ 0.5, 1, 2, and 3 $\mu$m. Points for $L =$ 0.5, 1 and 2 $\mu$m represent averages in a bin of 10 $\mu$m$^{-2}$. The inset is a plot of $\mathrm{log}\, \Delta$ as a function of $\mathrm{log}\, L$; the red line indicates a best-fit power exponent of $-1.0$. (c) The mean finite cluster size $S$ as a function of the density of nanoparticles (points represent averages in a bin of 2 $\mu$m$^{-2}$). The red dashed line is the theoretical expectation (described in text).\label{fig.5}}
\end{figure*}

Figure~\ref{fig.4} shows the density of wrinkles $\rho_{w}$ as a function of the nanoparticle density $\rho_{np}$ and the number of wrinkles per protrusion $\rho_{w}/\rho_{np}$ as a function of $\rho_{np}$ (inset). We find $\rho_{w}$ is almost zero below $\rho_{np} \approx 25$ $\mu$m$^{-2}$ (arrow b) and then begins to increase rather linearly with $\rho_{np}$ above  $\rho_{np} \approx 50$ $\mu$m$^{-2}$ (arrow c). We now analyze the behavior of the wrinkle density $\rho_{w}$ versus the nanoparticle density $\rho_{np}$ within a simple model.

We first consider nanoparticles placed at random on the substrate \cite{Supplementary}. Then wrinkles are placed with a probability $\Omega_{w}$ between neighboring nanoparticles separated by less than a cutoff length $X_c$. The probability $\Omega_{w}$ is expected to encompass not only the true probability to make a wrinkle between nanoparticles, but also all the information about wrinkle-orientation correlations, selecting only a fraction of all possible wrinkles. Since nanoparticles with more than three connected wrinkles are scarcely observed \cite{Supplementary}, we set three as the maximum number of wrinkles. Employing the probability density for a nanoparticle to have the $i$-th nearest nanoparticle ($i = 1, 2$, and $3$) at a distance $r$, $p_i(r)=2\left(\pi\rho_{np}\right)^{i+1}r^{2i+1}\mathrm{exp}\left(-\pi\rho_{np}r^2\right)/i!$ \cite{Gonzalez},
we find the density of wrinkles;
\begin{eqnarray}
\rho_{w}&=&\frac{\rho_{np}\Omega_w}{2}\sum^3_{i=1}\int^{X_c}_0dr\, p_i(r) \nonumber\\
&=&\frac{\rho_{np}\Omega_w}{2}\left[-\pi X_c^2 \rho_{np}\left(2+\frac{1}{2}\pi X_c^2 \rho_{np}\right)\mathrm{e}^{-\pi\rho_{np}X_c^2}\right.\nonumber\\
&&+\left.3\left(1- \mathrm{e}^{-\pi\rho_{np}X_c^2}\right)\right]. \label{prob}
\end{eqnarray}
The factor of $1/2$ in $\rho_w$ compensates for the double-counting of each wrinkle (i.e. from the particles at each end). In the small nanoparticle density limit $\rho_{np} \ll X_c^{-2}$, the density of wrinkles is $\rho_w = (1/2)\Omega_w\pi X_c^2\rho_{np}^2$, while in the large density limit $\rho_{np} \gg X_c^{-2}$, each nanoparticle has at least three neighboring nanoparticles within distance $X_c$, leading to $\rho_w = (3/2)\Omega_w\rho_{np}$. The red solid lines in Fig.~\ref{fig.4} are fits to Eq.~\ref{prob} with $\Omega_w = 0.54$ and $X_c = 120$ nm. The cutoff length is consistent with the observations. Furthermore, the agreement with $X_c$ predicted from the above elastic analysis is good. The model indicates a significant increase of the wrinkle density for the nanoparticle density larger than $(\pi X_c)^{-2}$, but also suggests that $\rho_w$ does not exhibit any singularity, i.e. wrinkling is a crossover phenomenon rather than a sharp transition.

\subsection{Percolation transition in the wrinkle network}

With increasing $\rho_{np}$, the connectivity of the wrinkle network increases, and we find a percolation transition at a threshold density $\rho_{c}$ (of order $X_c^{-2}$) at which the wrinkle network spans the system, as shown in the inset of Fig.~\ref{fig.5}a. The expansion of the network via wrinkling is a purely two-dimensional (2D) phenomenon. Thus, we analyze this behavior using 2D percolation theory \cite{Stauffer}: In Fig.~\ref{fig.5}a, we plot the probability $P$ that a given protrusion belongs to the percolating cluster spanning a region of size $L\times L$, where $L$ ranges from 1 to 3 $\mu$m. Also plotted is the prediction from 2D percolation theory: $P\sim(\rho_{np}-\rho_{c})^{\beta}$ for $\rho_{np} \geq \rho_{c}$ with $\rho_{c} =$ 87.5 $\mu$m$^{-2}$ as determined below and the ``magnetization'' exponent $\beta = 5/36$ \cite{Stauffer}, which reproduces the observations well. In Fig.~\ref{fig.5}b, we show the probability $\Pi$ that a cluster connects opposite sides of a region of size $L\times L$ ($L = 0.5, 1, 2$, and $3$ $\mu$m). For an infinite system, $\Pi = 1$ for $\rho_{np} \ge \rho_{c}$, while $\Pi = 0$ for $\rho_{np} < \rho_{c}$ \cite{Stauffer}. Indeed $\Pi$ displays a sharp transition around $\rho_{np} =$ 87.5 $\mu$m$^{-2}$ for $L = 3$ $\mu$m, indicating $\rho_{c}$ is in that vicinity.

Next, we probe the width $\Delta$ of the transition region, which is expected to scale as $L^{-1/\nu}$, where $\nu = 4/3$ is the correlation-length exponent \cite{Stauffer}. We define $\Delta$ as the difference in density for $\Pi = 0.9$ and $\Pi = 0.1$ in Fig.~\ref{fig.5}b. The inset of Fig.~\ref{fig.5}b shows that the data are well-fitted with $\nu = 1.0\pm 0.3$, consistent with the theoretical expectation.

Finally, we plot in Fig.~\ref{fig.5}c the mean size $S$ of the clusters (excluding the percolation cluster) as a function of $\rho_{np}$ with the theoretical prediction $S \sim |\rho_{np} - \rho_{c}|^{-\gamma}$, where $\gamma = 43/18$ is the ``susceptibility'' exponent \cite{Stauffer}. Some Monte Carlo simulations predict a much larger prefactor for $\rho_{np} \leq \rho_{c}$ (e.g., a critical amplitude ratio of $50 \pm 26$ for a continuum model \cite{Gawlinski}), in reasonable agreement with the observed ratio of $\sim 30$.

Thus, all measurements strongly support the existence of a 2D percolation transition at a critical nanoparticle density $\rho_{c} \approx$ 87.5 $\mu$m$^{-2}$. Since the only length scale is $X_c$, we obtain a universal number (i.e. independent of model parameters such as $\Gamma$, $E_{2D}$, or $d$) characterizing the wrinkle percolation transition: $\rho_{c}X_c^2 \approx 0.9$. In contrast, simple continuum percolation of penetrable discs of diameter $X_c$ leads to $\rho_{c}X_c^2 \approx 2.9$ \cite{Gawlinski}. This difference is a consequence of the unique structure of the wrinkle network (e.g., not more than three wrinkles merging at a given nanoparticle \cite{Supplementary}).

\subsection{Delamination of graphene multilayers}

\begin{figure*}[t]
\includegraphics[width=170 mm]{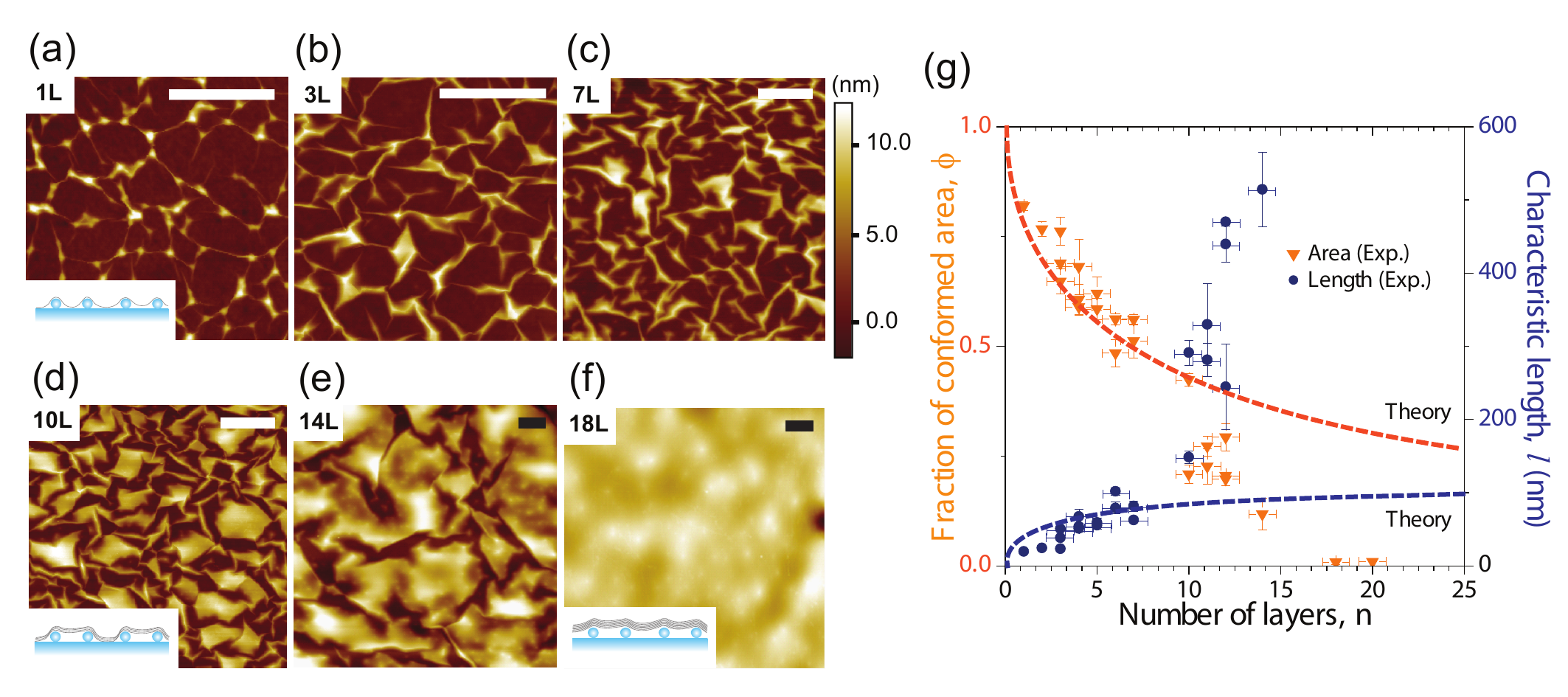}
\caption{Typical AFM images of (a) monolayer, (b) trilayer, (c) $7$-layer, (d) $10$-layer, (e) $14$-layer, and (f) $18$-layer graphene on SiO$_2$ with nanoparticle density $160\pm 24$ $\mu$m$^{-2}$. The scale bar in each image is 400 nm. The insets in (a), (d), and (f) are corresponding schematics of graphene films on nanoparticles to the AFM images. (g) The fractional area $\phi$ where graphene conforms to the substrate and the characteristic length $l$ of the delaminated domains, as functions of the number $n$ of graphene layers. The dashed curves are the theoretical expectations for $\phi$ (orange) and $l$ (blue) (described in text).\label{fig.6}}
\end{figure*}

Finally, we investigate morphological transitions which occur in multilayer graphene, using the same nanoparticle-templated substrates. The capability of multiple layers of graphene to mechanically screen an asperity recalls the use of multiple mattresses in an attempt to hide the presence of a pea in the fairy tale ``Princess and the Pea''. Figures~\ref{fig.6}a-f show typical AFM images of mono- and multi- layer graphene supported on nanoparticles of density $160 \pm 24$ $\mu$m$^{-2}$. The thicker graphene is partially suspended over the nanoparticles, as schematically shown in the insets of Figs.~\ref{fig.6}d and \ref{fig.6}f, with the delaminated area increasing with layer number $n$.

\begin{figure}[h]
\includegraphics[width=75 mm]{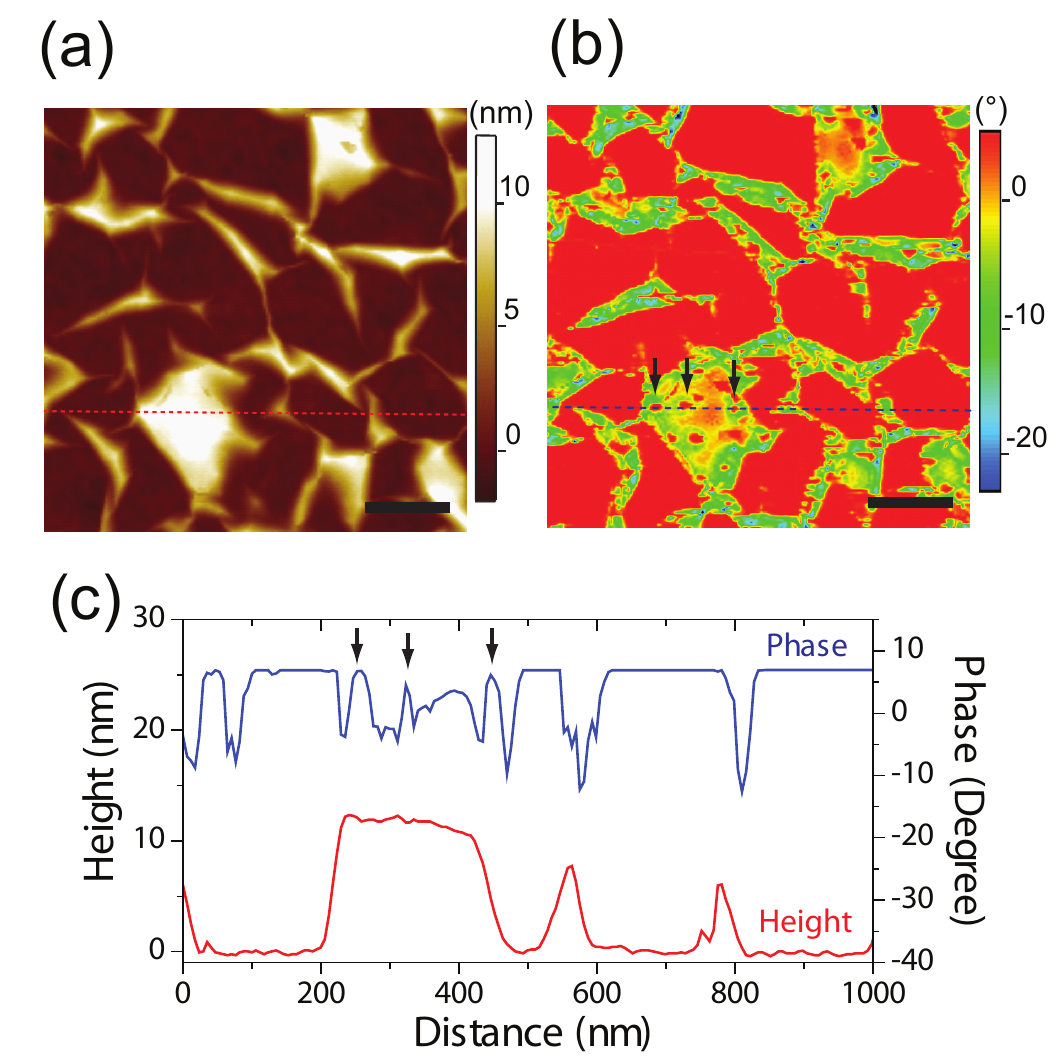}
\caption{Typical AFM (a) height and (b) phase images ($1\times 1$ $\mu$m$^2$) of 4-layer graphene supported on the nanoparticles. The scale bar is 200 nm. (c) Line profiles of the height and phase along the dashed red and blue lines shown in (a) and (b), respectively. The arrows correspond to those in (b), showing the locations of nanoparticles beneath graphene.\label{fig.7}}\end{figure}

The phase image in Fig.~\ref{fig.7} demonstrates that the suspended graphene is indeed supported by isolated nanoparticles; the mechanical response of the graphene to the AFM tip allows the detection of the hidden nanoparticle ``peas'' under the flat graphene ``mattress''. The phase image records the varying phase angle of the (oscillating) AFM cantilever as it interacts with an inhomogeneous sample surface. The phase angle increases with increasing local sample stiffness \cite{Magonov}. Figure~\ref{fig.7} shows that the phase image of 4-layer graphene discriminates between rigid supported regions (larger phase) and flexible suspended regions (smaller phase). The high, flat regions in the topograph show small, roughly circular regions of large phase indicating the locations of the nanoparticles (arrows) which support the surrounding suspended graphene (small phase).

Figure~\ref{fig.6}g shows the areal fraction $\phi$ of graphene in contact with the substrate and the characteristic length $l$ of the delaminated regions, as functions of $n$. As $n$ increases, a first transition occurs around $n = 10$, where $l$ increases rapidly (Fig.~\ref{fig.6}d; partial delamination); second, $\phi$ decreases and becomes negligibly small above $n \sim 15$ (Fig.~\ref{fig.6}f; complete delamination).

Surface-roughness--induced delamination of graphene has recently been studied theoretically \cite{Li1, Pierre-Louis, Kusminskiy} and experimentally \cite{Scharfenberg1, Scharfenberg2, Nicolle}. Models assume the elastic energy is dominated either by bending \cite{Pierre-Louis} or stretching \cite{Kusminskiy}. Here we consider each regime, and assume that the adhesion energy between SiO$_2$ and $n$-layer graphene $\Gamma_n$ is independent of $n$ for $n>1$ and has the value 1.9 eV/nm$^2$ \cite{Koenig}. In the bending-dominated model \cite{Pierre-Louis}, unbinding is controlled by a single dimensionless parameter $\alpha = (2\Gamma_n/\kappa_n)^{1/4}/[2\pi(\rho_{np}d)^{1/2}]$, where $\kappa_n$ is the bending rigidity of $n$-layer graphene for $n > 1$. Without interlayer sliding \cite{Poot}, continuum plate elasticity \cite{Landau} gives $\kappa_n = Et^3n^3/[12(1-\nu_g^2)]$, where $t = 0.335$ nm is the interlayer spacing, $E \approx 0.96$ TPa is the Young's modulus, and $\nu_g\approx 0.165$ is Poisson's ratio of monolayer graphene \cite{Lee}. The threshold for partial unbinding is predicted at $0.8\le\alpha\le1.3$, or $3\le n\le6$, with complete unbinding at $0.55\le\alpha\le0.75$, corresponding to $7\le n \le10$ \cite{Pierre-Louis}. Thus, the bending-dominated model underestimates the critical value of $n$ for unbinding, indicating that it overestimates the bending elastic energy. The one-dimensional character of the bending model limits its ability to make quantitative predictions. Furthermore, given the small radii of curvature in our experiment, the bending energy might well be reduced by partial interlayer sliding. Perfect sliding would give $\kappa_n = n\kappa$, leading to an unbinding threshold for $n$ a hundredfold larger; hence, interlayer sliding is extremely effective in relieving bending stress.

We then develop a stretching-dominated model, using Schwerin's solution for a membrane pushed by a point force \cite{Schwerin}. The solution gives the diameter of the detachment zone in $n$-layer graphene around a protrusion: $2R \approx d(4nE_{2D}/3\Gamma_n)^{1/4}$ (Refs.~\onlinecite{Kusminskiy} and \onlinecite{Zong}; see also Appendix C), where $nE_{2D}$ is the tensile rigidity of $n$-layer graphene. Then, we notice that the detached area around each protrusion is $\pi R^2$ and, furthermore, we assume that the detached areas produced by the wrinkles are negligible. Therefore, the typical length of the delaminated regions $l$ is estimated to be $2R$. The adhered area fraction $\phi$ is equivalent to the probability to have no nanoparticle in a domain of an area of $\pi R^2$, leading to $\phi = \mathrm{exp}(-\pi R^2\rho_{np})$ (see Appendix D for details). As shown in Fig.~\ref{fig.6}g, these predictions reproduce well the observed thickness dependence of $\phi$ and $l$ below $n \approx 10$, indicating that the stretching-dominated model for isolated protrusions accurately describes the small-$n$ limit where $\rho_{np} \ll l^{-2}$. However, $l$ increases and $\phi$ decreases much more rapidly than these predictions for $n > 10$, indicating collective effects become important. In order to understand the collective delamination in the high-nanoparticle-density regime $\rho_{np} > l^{-2}$, we may need to solve full elastic membrane equations, i.e., the F\"{o}ppl-von K\'{a}rm\'{a}n equations \cite{Landau} allowing for multiple nanoparticles.

\section{Pseudomagnetic fields in wrinkled graphene}

Finally, while a complete treatment is beyond the scope of this paper, we briefly discuss the potential impact of wrinkles on the electronic properties of graphene. Strain in graphene produces a vector potential in the electronic Hamiltonian, with strain gradients resulting in effective magnetic fields \cite{Castro Neto}. Such pseudomagnetic fields generated by strain gradients have been proposed as the basis of engineering graphene's electronic properties \cite{Levy, Guinea, Pereira2}.

We first evaluate the pseudomagnetic field generated by strain in the absence of wrinkling, corresponding to the case of small thickness or small nanoparticle density. In this case the elastic behavior of graphene on nanoparticles is predominantly determined by stretching, resulting in significant strain. The radial strain $\epsilon_r$ and the circumferential strain $\epsilon_{\varphi}$ both scale as $\sim \left(\Gamma d^2/E_{2D}\right)^{1/3}r^{-2/3}$ at $0 < r < R$, although the radial strain is more than 5 times as large as the circumferential strain (see Appendix C and Ref.~\onlinecite{Komaragiri}). The gauge fields induced by the axisymmetric strain can be written as $(A_r, A_{\varphi})  \approx (\Phi_0\beta/a)(\epsilon_r-\epsilon_{\varphi})(\cos 3\varphi, \sin 3\varphi)$, where $\Phi_0 = 10^{-15}$ Wb is the flux quantum, $\beta \approx 2$ is the change in the hopping amplitude between the neighboring atomic sites due to the lattice deformation \cite{Suzuura}, $a = 0.142$ nm is the lattice constant, and $\varphi$ is the azimuthal angle with $\varphi = 0$ in the zig-zag direction \cite{Guinea}. Then, the strain-induced pseudomagnetic field is given by $B_{\rm{eff}} = \partial_{\varphi}A_r/r - \partial_{r}A_{\varphi}-A_{\varphi}/r$. Thus, we find $B_{\rm{eff}} \sim (\Phi_0\beta /a)\left(\Gamma d^2/E_{2D}\right)^{1/3}r^{-5/3}\sin 3\varphi$. The strain in graphene on an isolated nanoparticle induces three-fold symmetric pseudomagnetic field profiles with maximum fields along the arm-chair directions. The pseudomagnetic field pattern is similar to a recent experiment \cite{Klimov}, in which suspended graphene was deformed by a sharp tip and pseudomagnetic fields were found to confine electrons to quantum dots with charging energies and level spacings both of order 10 meV. The divergence of the strain at $r = 0$ is cut off by the finite radius of the nanoparticles; thus, we may expect that maximum pseudomagnetic field appears at a radius comparable to the nanoparticle radius. We therefore estimate that the maximum pseudomagnetic field $B_{\rm{eff}}$ is of order 300 T for $d = 7.4$ nm and $r = d/2$, significantly greater than in Ref.~\onlinecite{Klimov}, suggesting that the impact on electronic properties may be even more profound.

Next we evaluate strain and strain-induced pseudomagnetic fields in a wrinkle. Strain along a wrinkle is given by $\epsilon_x \approx (\partial_x\zeta)^2$. Then, using Eq. \ref{eq.7}, we find the strain distribution $\epsilon_x \sim (\kappa/E_{2D})^{1/3}(X/2 \mp x)^{-2/3}$. The pseudomagnetic field is estimated to be $B_{\rm{eff}} \approx \Phi_0\beta\epsilon_x /(aW)$ \cite{Castro Neto, Kusminskiy}, where $W$ is the typical wrinkle width. In the strong adhesion limit $d(\Gamma/\kappa)^{1/2} \gg 1$, the wrinkle width $W$ can be estimated to be $\sim (\kappa/2\Gamma)^{1/2} \approx 1$ nm (see Appendix B). Thus, we find in the middle of a wrinkle the pseudomagnetic field has a broad minimum on the order of 10 T for $X = 100$ nm. 10 T is a large magnetic field compared to the disorder strength $1/\mu\sim 1$ T in typical graphene samples ($\mu$ being the electron mobility) and corresponds to an energy difference between 0th and 1st Landau levels of $\sim 1300$ K. Hence, we expect pseudomagnetic field effects due to wrinkles in graphene to be significant.

The pseudomagnetic field near particles in the wrinkled case will generally be more complicated, depending on the number of wrinkles terminating on the particle and their direction with respect to each other and the lattice. However, qualitatively we expect that since wrinkling reduces the in-plane strain around the nanoparticles, the pseudomagnetic field is also reduced. Recent molecular dynamics simulation results \cite{Neek-Amal} relevant to experiments by Tomori {\it et al}. \cite{Tomori} and our experiments of graphene on nanoparticles have indeed revealed that when nano-scale pillars supporting graphene are located far away from each other, graphene is detached only around the pillars and three-fold symmetric pseudomagnetic fields are induced around each pillar, while with decreasing distance between the pillars, graphene delaminates in regions between the pillars, resulting in complicated pseudomagnetic field profiles. Our observations of wrinkling and delamination combined with theoretical analysis based on a continuum elastic model can be used to place limits on strain distributions and, thus, on pseudomagnetic field maxima realizable in monolayer graphene through adhesion to patterned surfaces.


\section{Conclusions}

In conclusion, combining experiments with monolayer and multilayer graphene, we have obtained a global picture of the structural evolution of graphene membranes on surfaces of varying roughness. With increasing nanoparticle density (or graphene thickness), this evolution proceeds in five stages: conformal adhesion, wrinkling, wrinkle percolation, partial delamination, and complete delamination. The results can be used to place upper limits on the magnitude of pseudomagnetic fields generated in graphene by adhesion to patterned surfaces. Finally, the wrinkling and delamination are not specific to graphene; they should be a general feature of soft membranes adhered to rough surfaces, with implications for systems ranging from cell walls to fabrics.

\begin{acknowledgments}
This work was supported by the University of Maryland NSF-MRSEC under Grant No. DMR 05-20471 and NSF under Grant No. DMR 08-04976. The authors acknowledge E.~D. Williams for motivating our study of adhesion transitions in graphene. We thank Nissan Chemical America Corporation for providing samples of silica nanoparticle dispersions.
\end{acknowledgments}

\appendix

\section{Detailed geometric model for the wrinkle shape}

The stretching energy $E_s$, the bending energy $E_b$, and the adhesion energy $E_{\Gamma}$ given by Eqs.~\ref{eq.1}-\ref{eq.3} are reduced to 

\begin{eqnarray}
E_s&=&\frac{E_{2D}}{2}\int dx\, w(x)\epsilon_x^2\nonumber\\
&=&\frac{E_{2D}}{8}(\pi - \theta)\left[\frac{1}{\sin (\theta/2)}-1\right]^{-1}\int dx\, \zeta(\partial_x\zeta)^4 \nonumber\\ \label{eq.A1}
\end{eqnarray}
\noindent
\begin{eqnarray}
E_b&=&\frac{\kappa}{2}\int dx\, w(x)C_0(x)^2\nonumber\\
&=&\frac{\kappa}{2}(\pi - \theta)\left[\frac{1}{\sin (\theta/2)}-1\right]\int dx\, \zeta^{-1} \label{eq.A2}
\end{eqnarray}
\noindent
\begin{eqnarray}
E_{\Gamma}=\Gamma\int dx\, W = 2\Gamma Xd\tan(\theta/2). \label{eq.A3}
\end{eqnarray}

The combination between the bending and the adhesion at the foot of the wrinkle forces an equilibrium curvature at the contact line, $C_{eq} = (2\Gamma/\kappa)^{1/2}$ \cite{Seifert}. Then, assuming a simple geometry where a constant curvature region matches the constant slope region as shown in Fig.~\ref{fig.2}b, we find the bending energy $E_{b'}$ and the adhesion energy $E_{\Gamma'}$ of the curved zone:
\begin{equation}
E_{b'}=2\frac{\kappa}{2}\int dx\int dyC_{eq}^2 = X\left(\frac{\Gamma\kappa}{2}\right)^{1/2}(\pi-\theta) \label{eq.A4}
\end{equation}
\noindent
\begin{equation}
E_{\Gamma'}= 2\Gamma XC_{eq}^{-1}\tan\beta 
= X(2\Gamma\kappa)^{1/2}\tan\left(\frac{\pi-\theta}{4}\right),\label{eq.A5}
\end{equation}
where $2\beta$ is the angle of the curved region as shown in Fig.~\ref{fig.2}b.

Minimization of the total energy $E_{\rm{tot}} = E_s + E_b + E_{\Gamma}+E_{b'}+E_{\Gamma '}$ with respect to $\zeta$ leads to the following differential equation:
\begin{equation}
\zeta^2\left[3(\partial_x\zeta)^4\! +\! 12\zeta(\partial_x\zeta)^2\partial_{xx}\zeta\right]
+\frac{4\kappa}{E_{2D}}\left[\frac{1}{\sin(\theta/2)}\! -\! 1\right]^2\! =\! 0 \label{eq.A6}
\end{equation}
with the two boundary conditions $\zeta(\pm X/2) = 0$. We anticipate $\zeta$ to be symmetric with respect to $x$, so that $\partial_x\zeta$ should vanish at $x = 0$. However, Eq.~\ref{eq.A6} indicates that if $\partial_x\zeta$ vanishes at $x = 0$, either $\zeta$ or $\partial_{zz}\zeta$ should diverge. Since the solution with diverging $\zeta$ is physically inconceivable, we conclude that $\partial_{zz}\zeta$ should diverge. This indicates a discontinuity of the slope at $x = 0$. Physically this singularity would be regularized at small scales either by bending along the $x$ direction or by stretching along the $y$ direction. These contributions are expected to be small. As a result, we obtain a simple solution on both sides of the center of the wrinkle as described by Eq.~\ref{eq.7}.

\section{Scaling analysis for $X_c$}

Here we show that $X_c$ scales as $X_c \sim d(E_{2D}/\Gamma)^{1/4}$, analogous to scaling for the diameter detachment zones surrounding a local protuberance \cite{Kusminskiy}. We design a scaling analysis, neglecting $E_{\Gamma '}$ and $E_{b'}$. The total energy is of the form
\begin{equation}
E_{tot} = \kappa^{5/6}E_{2D}^{1/6}X^{1/3}[f_1(\theta)
+\Gamma d\kappa^{-5/6}E_{2D}^{-1/6}X^{2/3}f_2(\theta)],\label{eq.B1}
\end{equation}
where
\begin{eqnarray}
f_1(\theta) = \frac{4}{3^{1/2}}(\pi - \theta)\left[\frac{1}{\sin (\theta/2)}-1\right]^{2/3}\label{eq.B2}
\end{eqnarray}
and
\begin{eqnarray}
f_2(\theta) = 2\tan\frac{\theta}{2}.\label{eq.B3}
\end{eqnarray}
The minimization of Eq. \ref{eq.B1} with respect to $\theta$ leads to
\begin{eqnarray}
\partial_\theta f_1(\theta)+\Gamma d\kappa^{-5/6}E_{2D}^{-1/6}X^{2/3}\partial_\theta f_2(\theta) = 0\label{eq.B4}
\end{eqnarray}
and as a consequence
\begin{eqnarray}
\theta = f_3(\Gamma d\kappa^{-5/6}E_{2D}^{-1/6}X^{2/3}).\label{eq.B5}
\end{eqnarray}
The critical spacing $X_c$ is given by a condition that $\zeta(0) = d$. Hence, using Eq. \ref{eq.7}, we find
\begin{eqnarray}
X_c&=&d^{3/2}\frac{2^{3/2}}{3^{3/4}}\left(\frac{E_{2D}}{\kappa}\right)^{1/4}\left[\frac{1}{\sin (\theta (X_c)/2)}-1\right]^{-1/2} \nonumber \\
&=&d^{3/2}\left(\frac{E_{2D}}{\kappa}\right)^{1/4}f_4(\theta) \nonumber \\
&=&d^{3/2}\left(\frac{E_{2D}}{\kappa}\right)^{1/4}f_4(f_3(\Gamma d\kappa^{-5/6}E_{2D}^{-1/6}X^{2/3})), \nonumber\\ \label{eq.B6}
\end{eqnarray}
where
\begin{eqnarray}
f_4(\theta) = \left[\frac{1}{\sin (\theta /2)}-1\right]^{-1/2}.\label{eq.B7}
\end{eqnarray}
One can check by substitution that
\begin{eqnarray}
X_c = d^{3/2}\left(\frac{E_{2D}}{\kappa}\right)^{1/4}f_5(\Gamma d^2/\kappa)\label{eq.A8}
\end{eqnarray}
with $f_6(u) = f_5(uf_6(u)^{2/3})$ and $f_5(u) = f_4(f_3(u))$. We now define the elastic thickness $h_{el} = (\kappa/E_{2D})^{1/2}$ and the equilibrium contact curvature $C_{eq} = (2\Gamma/\kappa)^{1/2}$. Letting $f_7(u) = f_6(u^2/2)$, we rewrite $X_c$ as
\begin{eqnarray}
X_c = \frac{d^{3/2}}{h_{el}^{1/2}}f_7(C_{eq}d),\label{eq.A9}
\end{eqnarray}
which is the general scaling form of the solution.

We now consider two asymptotic limits; the strong adhesion limit $C_{eq}d \gg 1$ and the weak adhesion limit $C_{eq}d \ll 1$. In the strong adhesion limit, the opening angle of the wrinkle $\theta$ goes to zero. Then, one has $f_1(\theta) \sim \theta^{-2/3}$ and $f_2(\theta) \sim \theta$. Hence $f_3(u) \sim u^{-3/5}$. Since $f_4(\theta) \sim \theta^{1/2}$, one has $f_5(u) = f_4(f_3(u)) \sim [f_3(u)]^{1/2} \sim u^{-3/10}$, and finally $f_6(u) \sim u^{-1/4}$. Therefore one has:
\begin{eqnarray}
X_c \sim \frac{d^{3/2}}{h_{el}^{1/2}}(C_{eq}d)^{-1/2}\sim d(E_{2D}/\Gamma)^{1/4}\label{eq.B10}
\end{eqnarray}
and
\begin{eqnarray}
\theta \sim (C_{eq}d)^{-1},\label{eq.B11}
\end{eqnarray}
which also confirms that the small-$\theta$ limit corresponds to the large-$C_{eq}d$ limit (strong adhesion limit).

 Alternatively, setting $\Theta = \pi - \theta$, we redo the above scaling analysis in the weak adhesion limit $C_{eq}d \ll 1$ with the argument of $f_1$, $f_2$, and $f_4$ being $\Theta$ instead of $\theta$. Then $f_1(\Theta) \sim \Theta^{7/3}$, and $f_2(\Theta) \sim \Theta^{-1}$, so that $f_3(u) \sim u^{3/10}$. Also $f_4(\Theta) \sim \Theta^{-1}$, leading to $f_5(u) \sim u^{-3/10}$ and $f_6(u) \sim u^{-1/4}$. Consequently, we obtain once again
\begin{eqnarray}
X_c \sim \frac{d^{3/2}}{h_{el}^{1/2}}(C_{eq}d)^{-1/2}\sim d(E_{2D}/\Gamma)^{1/4}\label{eq.B12}
\end{eqnarray}
and
\begin{eqnarray}
\Theta \sim (C_{eq}d)^{1/2}.\label{eq.B13}
\end{eqnarray}
This solution is consistent with the weak adhesion limit because $\Theta \ll 1$ implies $C_{eq}d \ll 1$.

\section{Stretching of graphene on a nanoparticle}

\begin{figure}[h]
\includegraphics[width=75 mm]{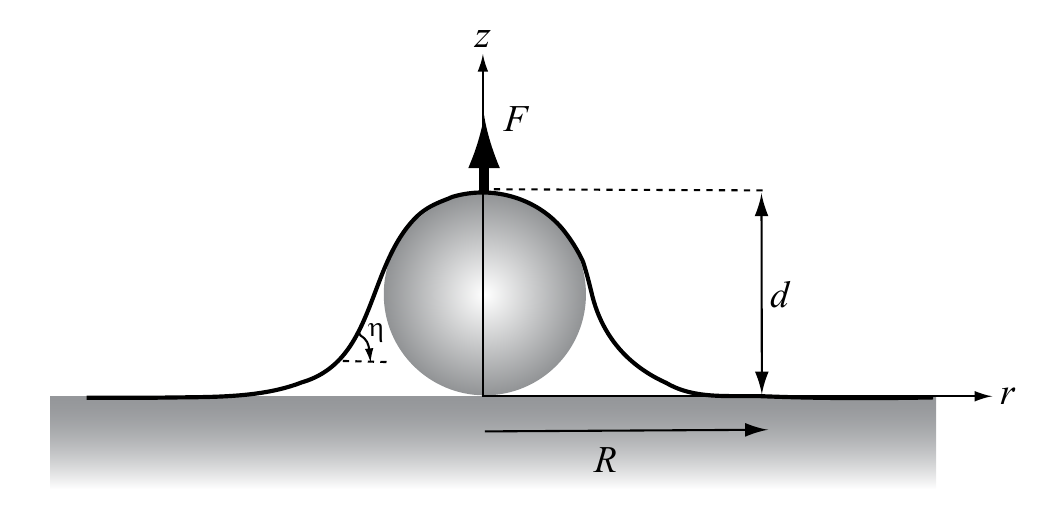}
\caption{Schematic of graphene supported on a single nanoparticle. The detachment length is $R$.}\label{fig.8}
\end{figure}

The typical extent of the detachment zone of graphene caused by the presence of a nanoparticle is discussed here. In the simplest model, we use the power-law solution of Schwerin \cite{Schwerin}, in the regime dominated by stretching. In this regime, the bending rigidity only contributes as boundary layer effects at the attachment lines \cite{Komaragiri}. Consequently, the Schwerin solution does not match tangentially to the substrate and the nanoparticle (the matching being forced at a smaller scale controlled by bending rigidity). Assuming that the nanoparticle diameter $d$ is much smaller than the radius $R$ of the detachment zone (see Fig.~\ref{fig.8}), we obtain the angle of rotation $\eta (r) = [8F/(9\pi E_{2D}r)]^{1/3}$ and the vertical distance from the substrate $Z = [3R^2F/(\pi E_{2D})]^{1/3}$ (see Fig.~\ref{fig.8} and Ref.~\onlinecite{Komaragiri}), where $F$ is the force on the apex. However, the above solution does not match the boundary conditions, and a better approximate numerical solution is: $Z = g(\nu_g)(R^2F/E_{2D})^{1/3}$, where $g(\nu_g) = 1.0491-0.1462\nu_g -0.15827\nu_g^2$ \cite{Komaragiri}. For graphene, $\nu_g = 0.165$ \cite{Lee}, and $g(\nu_g) \approx 1.029$ is very close to $(3/\pi)^{1/3} \approx 0.984$ so that we can safely use the direct Schwerin solution.

The elastic stretching energy can be calculated from a gedanken experiment, where the height $Z$ is increased with constant $R$:
\begin{eqnarray}
E(Z) &=& \int^Z_0dz F(z) \nonumber \\
&=& \frac{\pi E_{2D}}{3R^2}\int^Z_0dz z^3 \nonumber \\
&=& \frac{\pi E_{2D}Z^4}{12R^2}.\label{eq.D1}
\end{eqnarray}
Assuming that the apex height is equal to the diameter $d$ of the nanoparticle, one has $Z = d$, and the total energy reads
\begin{eqnarray}
E(d)=\frac{\pi E_{2D}d^4}{12R^2}+\Gamma\pi R^2.\label{eq.D2}
\end{eqnarray}
Minimizing the total energy with respect to $R$, one finds $2R/d = (4E_{2D}/3\Gamma)^{1/4}$ as suggested in Ref.~\onlinecite{Kusminskiy}.

Assuming that a pre-existing strain is negligibly small, one can find radial and circumference components of strain present in an isolated graphene protrusion at $0 < r < R$:
\begin{eqnarray}
\epsilon_{r}(r)&=& \frac{1}{E_{2D}}\left(\frac{G}{r}\cos\eta + H\sin\eta - \nu_g \frac{dG}{dr}\right) \nonumber \\
&=& \frac{3-\nu_g}{4}\left(\frac{4\Gamma}{9E_{2D}}\right)^{1/3}\left(\frac{d}{r}\right)^{2/3} \\
\epsilon_{\varphi}(r) &=& \frac{1}{E_{2D}}\left(\frac{dG}{dr} - \nu_g\frac{G}{r}\cos\eta - \nu_g H\sin\eta\right) \nonumber \\
&=& \frac{1-3\nu_g}{4}\left(\frac{4\Gamma}{9E_{2D}}\right)^{1/3}\left(\frac{d}{r}\right)^{2/3},
\end{eqnarray}
where $G = F/(2\pi\eta) = 9E_{2D}\eta^2r/16$ is the stress variable in the radial direction, $H = F/(2\pi r)= 9E_{2D}\eta^3/16$ is the vertical sheet stress, and we assume $\eta \approx 0$ at $0 < r < R$ \cite{Komaragiri}.


\section{Area probability distribution}

The probability $Q(A)$ that a nanoparticle has no nanoparticles in a domain of area $A$ obeys $Q(A+dA) = Q(A) - dA\,\rho_{np}Q(A)$, leading to $dQ(A)/dA = -\rho_{np}Q(A)$. Using the normalization constraint
\begin{eqnarray}
\lim_{A \to 0} Q(A) = 1,\label{eq.E1}
\end{eqnarray}
we find $Q(A) = \mathrm{exp}(-\rho_{np}A)$.

\end{document}